# Determine Disturbance Locations in Power Grids using Bicubic 2D Interpolation on Electromechanical Wave-front Propagation Delay

Shutang You

**Abstract:** This study presents a method to locate power system disturbance using wide-area synchrophasor measurements. The merits of the proposed method include robustness and easy for visualization. In addition, the proposed method facilitates the calculation of electromechanical wave propagation speed distribution. An example of locating the disturbance and generating the propagation speed distribution is demonstrated based on FNET/GridEye, a distribution-level wide-area measurement system. Without losing generality, the proposed method can be implemented in any other wide-area measurement systems.

## 1. Introduction

As renewable generation increases, due to system uncertainty and complexity, modern power system operation increasingly relies on high-resolution real-time situation awareness systems. Wide-area synchrophasor measurement systems (WAMS) can provide measurements with a much higher time resolution than conventional measurement systems. Therefore, WAMS is considered as the state-of-the-art technology to monitor power grid dynamics [1, 2]. The utility-level sensor in WAMS is called Phasor Measurement Units (PMU). PMUs are expensive, installed in high voltage environments, and need complex installation and maintenance procedures, which hinder their deployment substantially. Also, PMU measurement is considered as property data, which makes it difficult for exchanging between utility companies.

As the WAMS deployed at the distribution level, FNET/GridEye has been monitoring power grids using synchrophasor technology for more than 10 years. Measured data include frequencies, voltage phasor angles and magnitudes, as well as power quality data. FNET/GridEye consists of two major parts: over 300 frequency disturbance recorders (FDRs) as the sensors installed on world-wide power grids, and the data centre hosted at the University of Tennessee, Knoxville (UTK) and Oak Ridge National Laboratory (ORNL). As a quickly-deployable distribution-level synchrophasor measurement system, multiple applications have been developed. Compared with WAMSs owned by utilities, FNET/GridEye has many unique applications. One of such applications is disturbance location determination.

Power grids are subjected to various types of disturbances frequently, such as load variations, generator and line trip, and faults. Many disturbances are minor but large disturbances or events may lead to system emergency states. Automatic event detection and locating events can improve situation awareness and activate appropriate controls to prevent cascading failures. Taking the generation trip location as an example, detecting a generation trip event will allow demand side frequency response. Further, locating a

Email: syou3 @utk.edu.

generation trip will allow each balancing authority to minimize frequency excursions and constrain the impact of the event through proper frequency control strategies, such as activating system reserve and leveraging responsive loads.

The type of event can be identified based on the footprints on the frequency or angle profiles. In the FNET/GridEye system, different types of events will have unique characteristics on the frequency measurements. For example, typical load shedding will cause a sharp increase on the frequency profile, while generator disconnection will result to a sharp frequency decrease. Line trip events will cause oscillations on frequency but without change on the average frequency. A fault event will cause a relatively local frequency drop. These features can be used to detect different types of events, as investigated in [3, 4].

This paper mainly focuses on the event location problem. Existing disturbance locating methods have two major steps: obtaining the FDR response time and estimating the location. The FDR response time can be obtained using the time of the frequency measurement passing a threshold. For estimating the disturbance location, the least-square disturbance location approach needs the assumption that the electromechanical wave propagation speed is already known and keeps constant throughout the system [5]. Otherwise, the method will give a series of suspicious locations: one location for each propagation speed. However, in practice, there does not exist a uniform speed that is always valid for one power grid. Actually, the propagation speed varies with the system conditions such as unit commitment and load dynamics, making the approach difficult to be applied. The method described in [6] combines power grid models and measurements to locate generator trips. However, this method heavily relies on the power grid model in calculating the propagation distance. This model-dependency makes it not applicable when the model is not available or the system topology changes with operation conditions. Ref. [7] described a non-parametric approach. This approach determines the likelihood of event location by partition the plane into two parts for each two measurements. The result of this approach may give large and irregular areas as suspicious locations of a disturbance and it is sometimes hard to choose the probability distribution function.

In summary, existing disturbance location approaches either relies on system models or are highly depend on parameters [8]. This paper presents a robust and parameter-insensitive method to locate disturbance location based on FNET/GridEye. To pinpoint the event location, this method combines the Delaunay triangulation [9, 10] and the bicubic 2D interpolation [11] by reconstructing the wave arrival time. In addition, under the framework of the proposed method, it is convenient to calculate the disturbance propagation speed distribution in the power grid, which is valuable information to ensure reliable protection actions under high renewable penetrations [12-14].

The rest of this paper is organized as follows: Sections 2 gives an overview on the FNET/GridEye system. Section 3 describes the new event location method and its implementation in FNET/GridEye. Conclusions are presented in Section 4.

## 2. Overview On FNET/GridEye, A Distribution-Level Wide-Area Measurement System

### *2.1. Frequency Disturbance Recorder*

The idea of distribution-level synchrophasor technology makes it possible to significantly decrease WAMS costs and simply deployment [15]. Nevertheless, special technical challenges arise when designing hardware and software for synchrophasor measurement sensors in the distribution system [16]. For example, different from transmission systems, distribution systems have much worse power quality due to the harmonics and distortions produced by various electric appliances. Under these circumstances, the distribution-level sensors should be capable to capture power grid dynamics at noisy system ends.

Embedded with a microprocessor for sampling and estimating frequency and voltage phasor, as well as other modules for GPS time synchronization and Ethernet communication, FDR features low manufacturing cost, which is about one tenth of a typical PMU [17]. Besides, FDR simplifies installation procedures to plug-and-play. FDR does not sacrifice its accuracy for low cost and quick deployment [18]. In fact, FDR has comparable or even higher accuracy than its counterparts. For example, the target of Micro-FDR is ±0.05° in angle measurement accuracy, which is surpassed by FDR [19]. So far, three generations of FDRs have been developed for improved measurement accuracy and data quality. Fig.1 shows the most widely deployed Generation-II FDR. Updates on Generation-III FDR include enhanced functionalities on power quality measurements, and more importantly, accuracy improvement archived by hardware and algorithm advancement. Its measurement accuracy reaches a record of ±0.00006 Hz (for frequency) and ±0.003° (for voltage angle) under steady states.

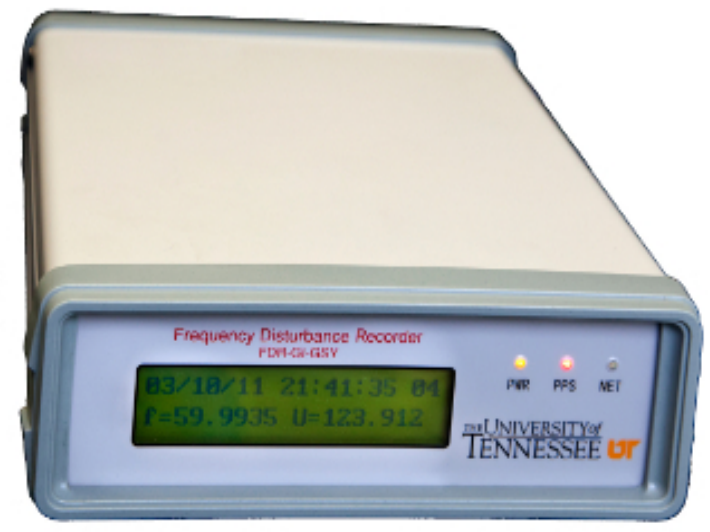

***Fig. 1.*** *Generation-II FDR*

The deployment of FDR in North America and its FNET/GridEye world-wide coverage by 2015 are shown in Fig.2 and Fig.3, respectively. Real-time phasor measurements collected by FDRs at the indicated locations are transmitted via Ethernet and collected by the data centre hosted at UTK and ORNL. These measurements enable performing multiple functionalities including online monitoring, online analysis, and offline data mining, revealing valuable insights into power grid dynamic behaviours.

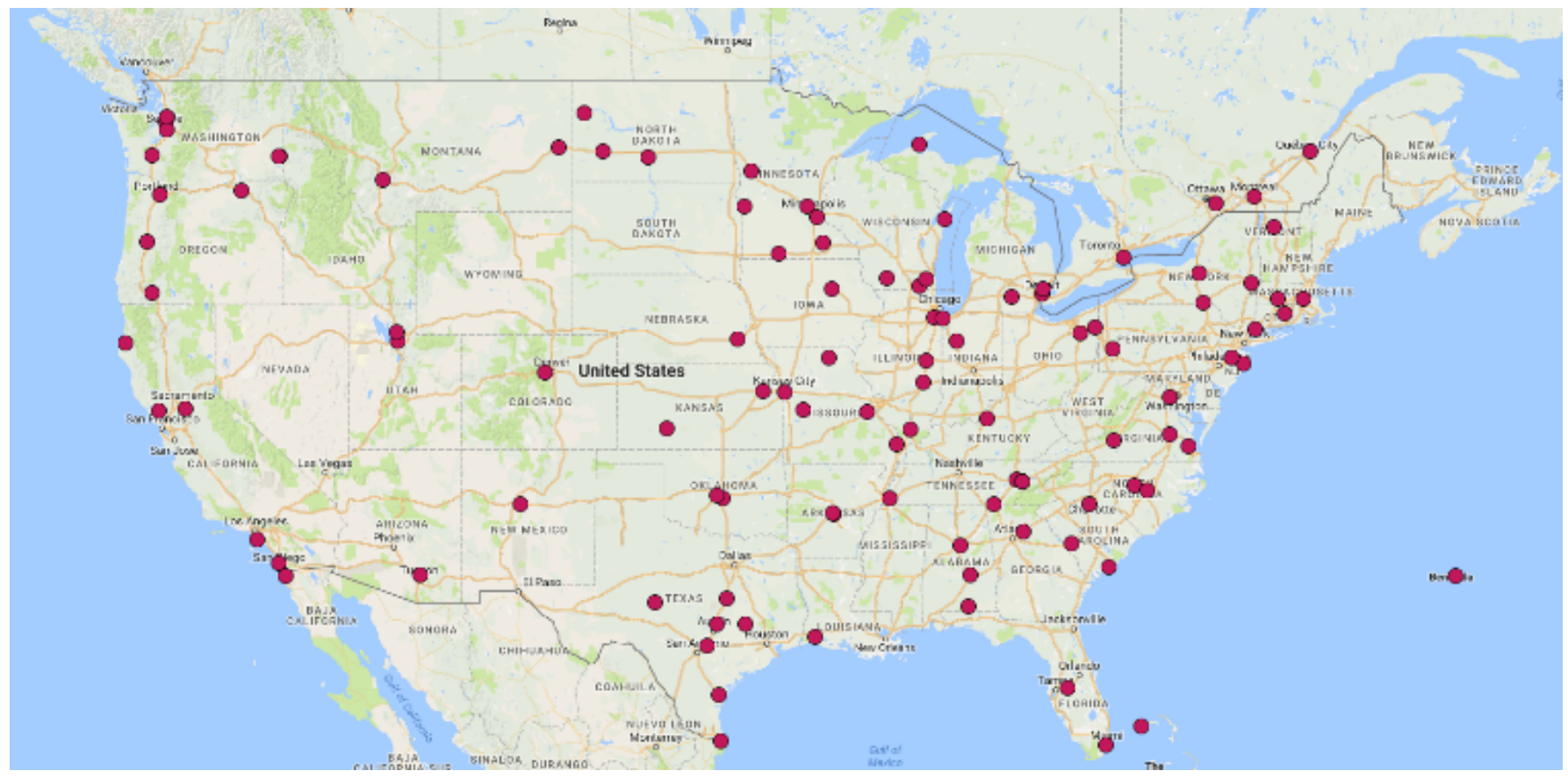

*Fig. 2. The FDR location map in North America*

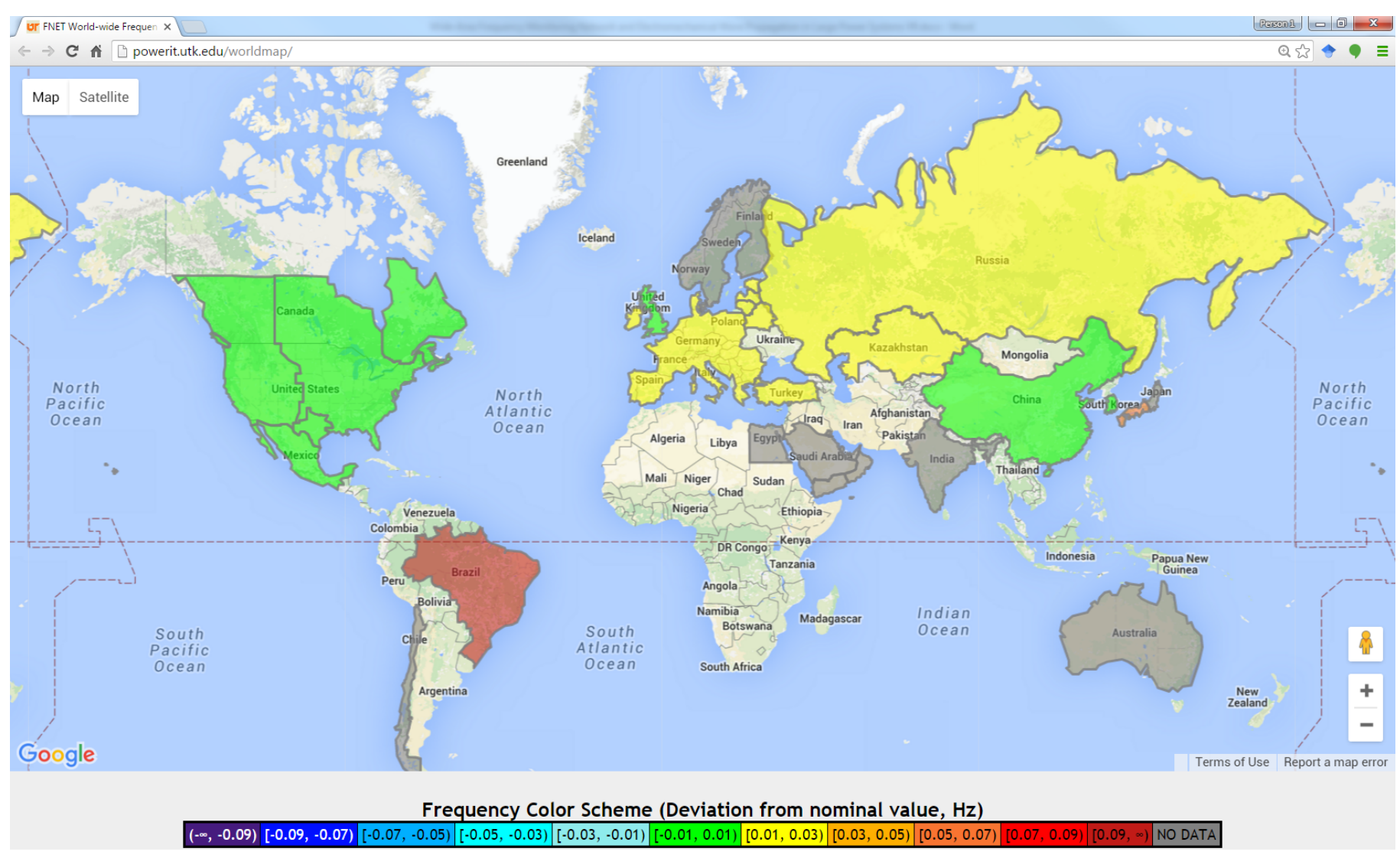

***Fig. 3.** World-wide FDR deployment and the frequency map*

### 2.2. FNET/GridEye Data Center

The FNET/GridEye data center is capable of managing, technically processing, and safely archiving the measurements in a systematic way. It has a multi-layer structure as shown in Fig.4 [18].

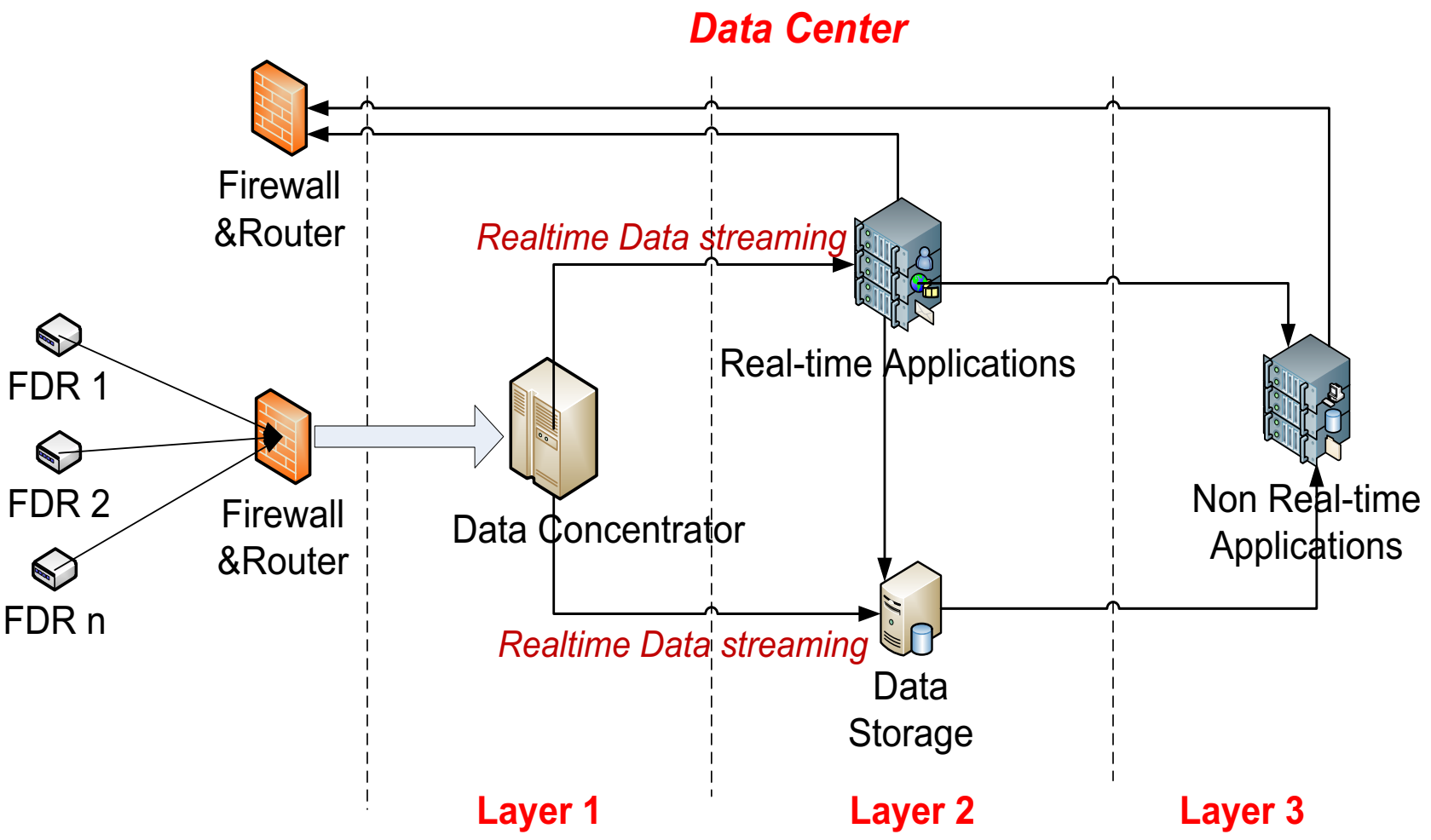

***Fig. 4.*** *The FNET/GridEye data centre structure*

The first layer of the data centre is the data concentrator, in which TCP/IP data packages are extracted, interpreted, error-checked, time-aligned, and then streamed into the subsequent layer. The second layer includes two agents: the real-time application agent and the data storage agent. Various real-time application modules are running on the real-time application agent to monitor power grid dynamic behaviours based on streaming data. For example, on this agent, the real-time event detection module monitor disturbances on the interconnection scale and sends out event alerts to system operators. Meanwhile, the data storage agent archives phasor measurement data streams and outputs from the real-time application agent for off-line applications. All data are archived in an efficient format to preserve data integrity while saving space. In the third layer, the non-real-time application agent runs offline applications to further exploit the archived data [20]. The archived data is a valuable information source for power grid research. For example, the dynamic models of US power grids could be validated through comparing the actual system frequency responses (stored in the archived phasor measurements) with the model-based simulation results [21].

The multi-layer structure of the FNET/GridEye server facilitates efficient concentrating, processing, and archiving wide-area measurements so as to successfully meet the timeliness requirements of various functionalities [22, 23]. Based on the FNET/GridEye platform, a variety of visualization and analytics applications have been developed, and they are widely adopted by the academia, the industry, and government agencies [24]. These applications enable system operators to keep better aware of the spatiotemporal evolvement of power grid dynamics rendered by various disturbances and changing environments.

## 3. Disturbance Location Determination Based on Electromechanical Wave Propagation

Since the interconnection-level power grid is large and the FDR distribution is coarse, the FDR measurement in disturbance location is not available except for those cases in which disturbances happen in FDR-installed locations. A disturbance of the power grid results to speed changes of generator rotors, similar to the phenomenon of wave dissemination on a water surface. The speed of generator rotors, which is denoted by the frequency measurement, is a good indicator of the electromechanical wave impact. To determine the disturbance location, the proposed method uses frequency measurements of FDRs sparsely distributed in a wide area at the distribution level. Besides, this method adopts Delaunay triangulation and bicubic 2D interpolation to locate disturbance in a more accurate way. The workflow of the proposed method includes the following steps:

1) align the frequency measurement based on the GPS timestamp;
2) filter and interpolate the frequency measurement to eliminate aliasing and reconstruct frequency profile;
3) extract the relative arrival time;
4) perform Delaunay triangulation of FDR GPS coordinates;
5) interpolate the response time in the spatial domain using bicubic 2D interpolation;
6) locate the disturbance by searching the minimum relative arrival time;
7) Validate measurements to eliminate impact from bad data; and
8) calculate the propagation speed distribution.

The structure of the workflow is shown in Fig. 5.

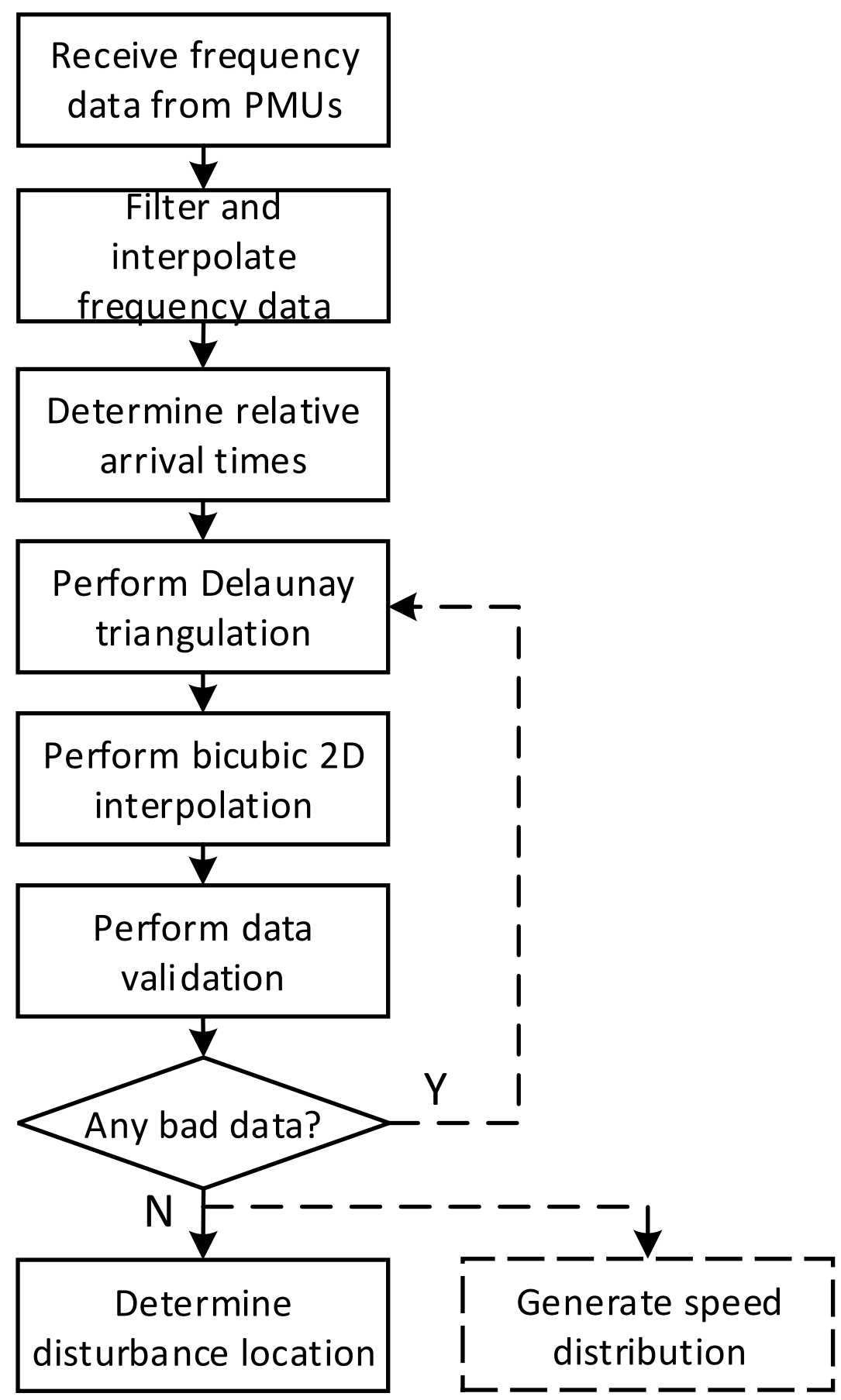

***Fig. 5.*** *The flow chart of the disturbance location method*

### *3.1. Frequency measurement filtering, interpolation, and relative arrival time*

FDRs measure the frequency at different locations during a power grid disturbance. The frequency data during a disturbance event is filtered with a moving average filter to remove the high frequency noise. Thus, an equivalent form of the average filter for frequency measurement is defined as:

$$\bar{f} = \frac{1}{N}\sum_{i=1}^{N} f_i \tag{1}$$

where *N* is the size of the moving window. For the reporting rate of ten measurements per second in the current FNET/GridEye system, *N* is set to 5. After filtering, the frequency measurements are interpolated before obtained the time delay of arrival for each FDR. The aim of implementing interpolation is to reconstruct the frequency during the periods between the reporting time snapshot, so that a better estimation of the arrival time can be obtained. In this paper, linear interpolation is applied.

As an example, a generation trip disturbance happened at 13:48:05 (UTC) on Nov. 21$^{st}$, 2014. During this event, 61 FDRs were streaming data at different locations of the U.S. Eastern Interconnection (EI) power grid. The frequency dropped from 60.005Hz to around 59.975Hz through an “L” profile, which is a typical frequency response of a generation trip disturbance. Fig. 6 shows the filtered and interpolated frequency data of the detected disturbance event from multiple FDR units. The start time of the decrease of frequency relates to the distance of the FDR and the disturbance location. Those FDR units that are closer to the disturbance source have sharper and early decrease in frequency measurement. To calculate the relative arrival time, a threshold of frequency $f_T$ is applied. Subroutines are developed to detect the disturbance, determined the threshold $f_T$ and the common reference time $t_R$ for a specific disturbance automatically. Determining the threshold value $f_T$ involves three steps.

1) Calculate the system average frequency;

2) Calculate the ROCOF (rate of change of frequency) of the average frequency;

3) Determine the event start time based on ROCOF; and

4) $f_T = \text{frequency at the event start time} - \Delta f$ .

$\Delta f$ is the frequency deviation threshold. This value can be easily determined by examining typical disturbances in a specific system. For EI, this value is 0.005Hz. In step 3, the event start time is determined by confirming that after a timestamp (the event start time), the majority ROCOF values (75% for EI) during a consecutive period (4 second for EI) are larger than a threshold ROCOF (1mHz/s for EI). A majority (75% for EI) of ROCOF passing the threshold is enough to confirm the event occurrence because of the influence of oscillations stirred up by event disturbances. The relative arrival time is then defined as the difference between the common reference time $t_R$ and the time of a FDR's frequency exceeding the threshold. For the case shown in Fig. 6, $f_T$ is calculated to be 60.0014Hz and $t_R$ is selected as18.8s, respectively, as shown in Fig. 7. It is worth noting that the purpose of introducing $t_R$ is to set a common reference of the response time. It does not influence the event location result.

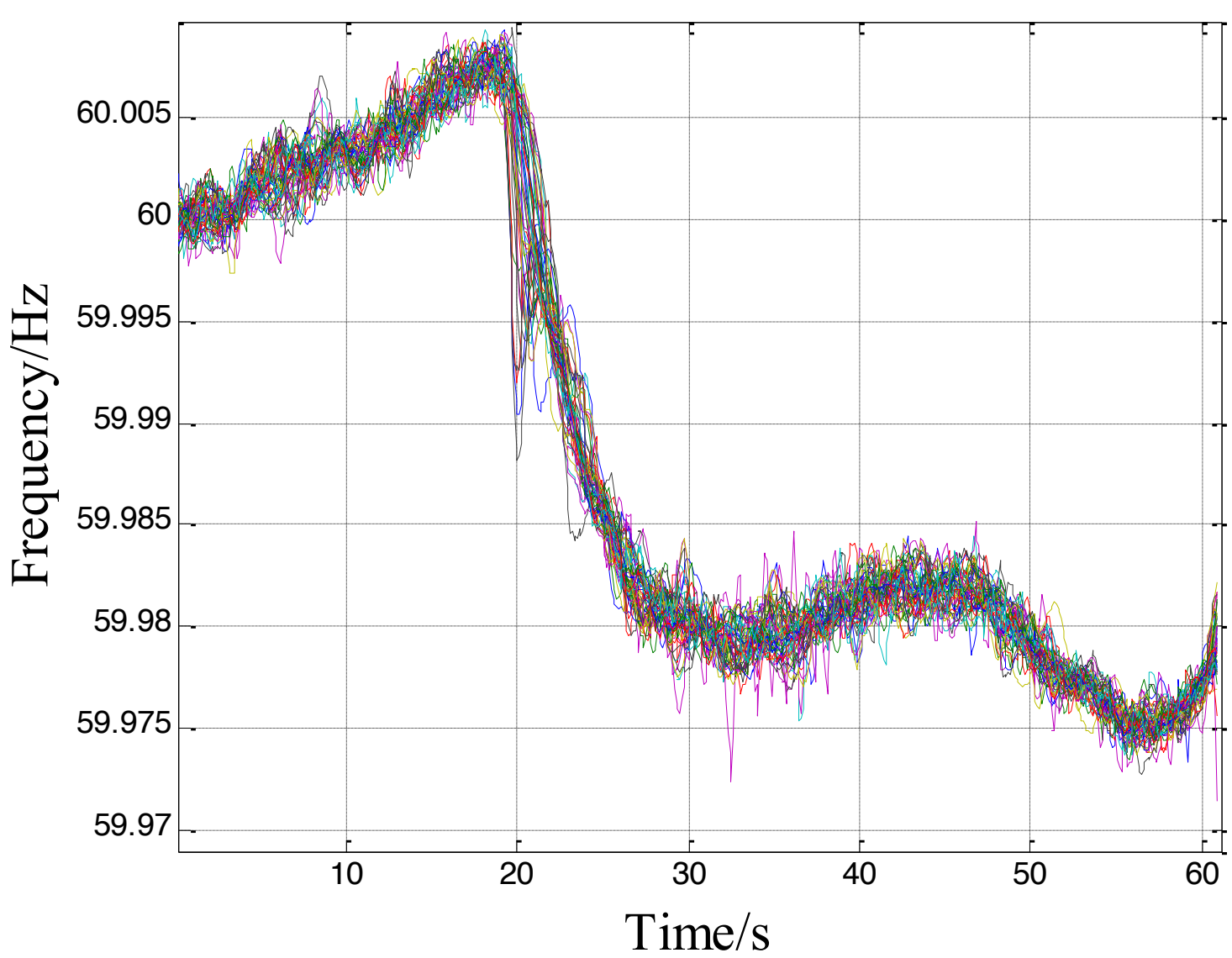

*Fig. 6.* *Filtered frequency (5 points median) of the detected disturbance*

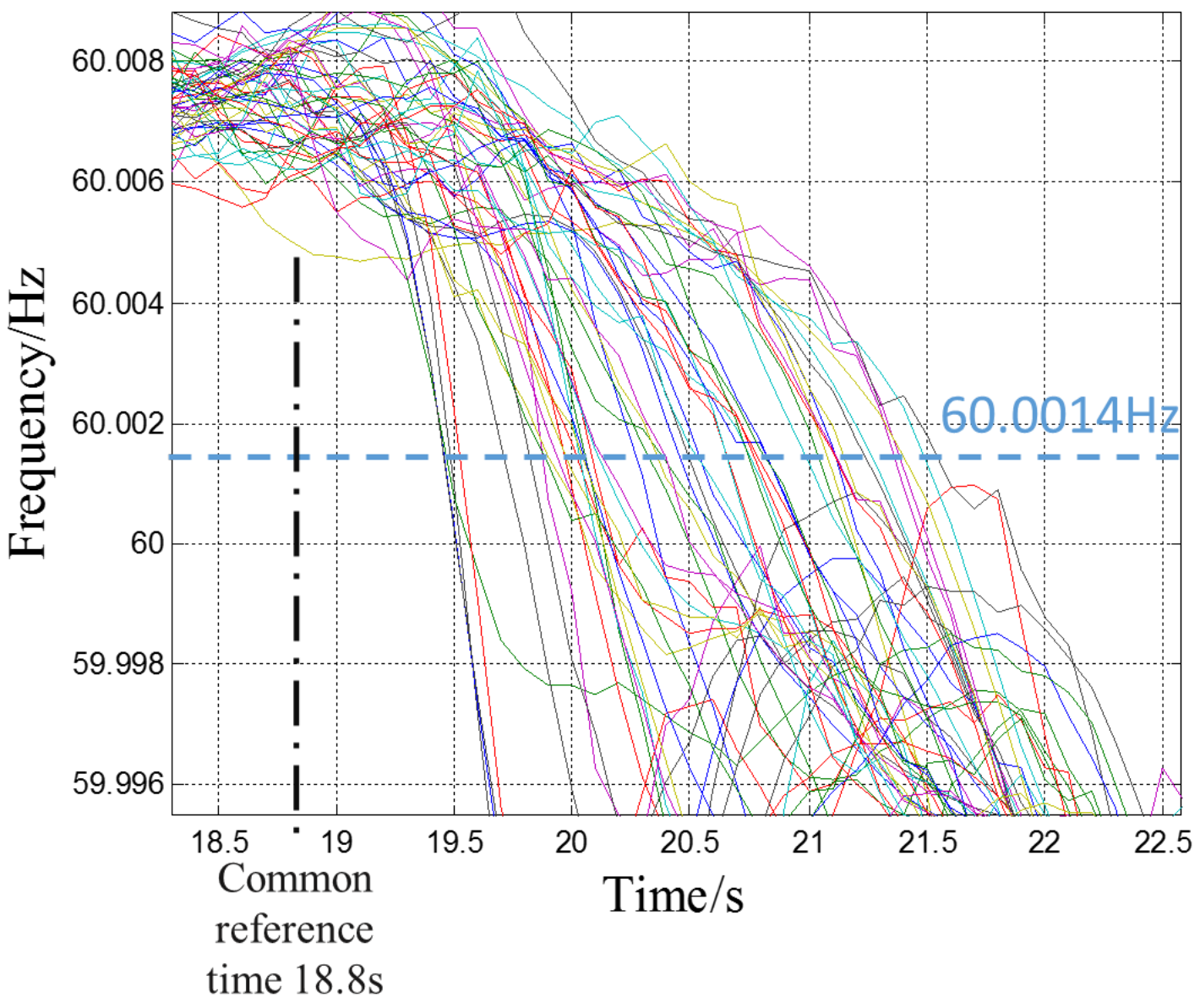

*Fig. 7.* *Relative arrival time*

Table 1 lists the relative arrival time of some FDRs and Fig. 8 graphically shows the relative arrival time of FDRs at different locations. Blue dots represent earlier response time whereas red ones represent relatively later responses. It shows that FDRs in four states: Kansas, Missouri, Arkansas, and Louisiana, had the smallest response time recorded (~0.7s). The New England area, which was remote from the disturbance, had the largest delay of response.

**Table 1.** Relative arrival time of some FDRs

| FDR # | FDR Location | | Relative Arrival Time (s) |
|---|---|---|---|
| | State | Location (City or Company) | |
| 844 | KS | Dodge City | 0.7870 |
| 941 | KS | Wakeeney | 0.7608 |
| 647 | AR | Little Rock | 0.7709 |
| 979 | LA | Shreveport | 0.8273 |
| 886 | MO | Kansas City | 0.7595 |
| 777 | NE | LES | 0.9996 |
| 756 | MO | Franklin | 1.1798 |
| 1027 | MS | Jackson | 1.1941 |
| 906 | TX | Pleasant Hill | 1.1252 |
| 792 | VA | Hay Market | 2.1406 |
| 1048 | NY | Fredonia | 2.2970 |

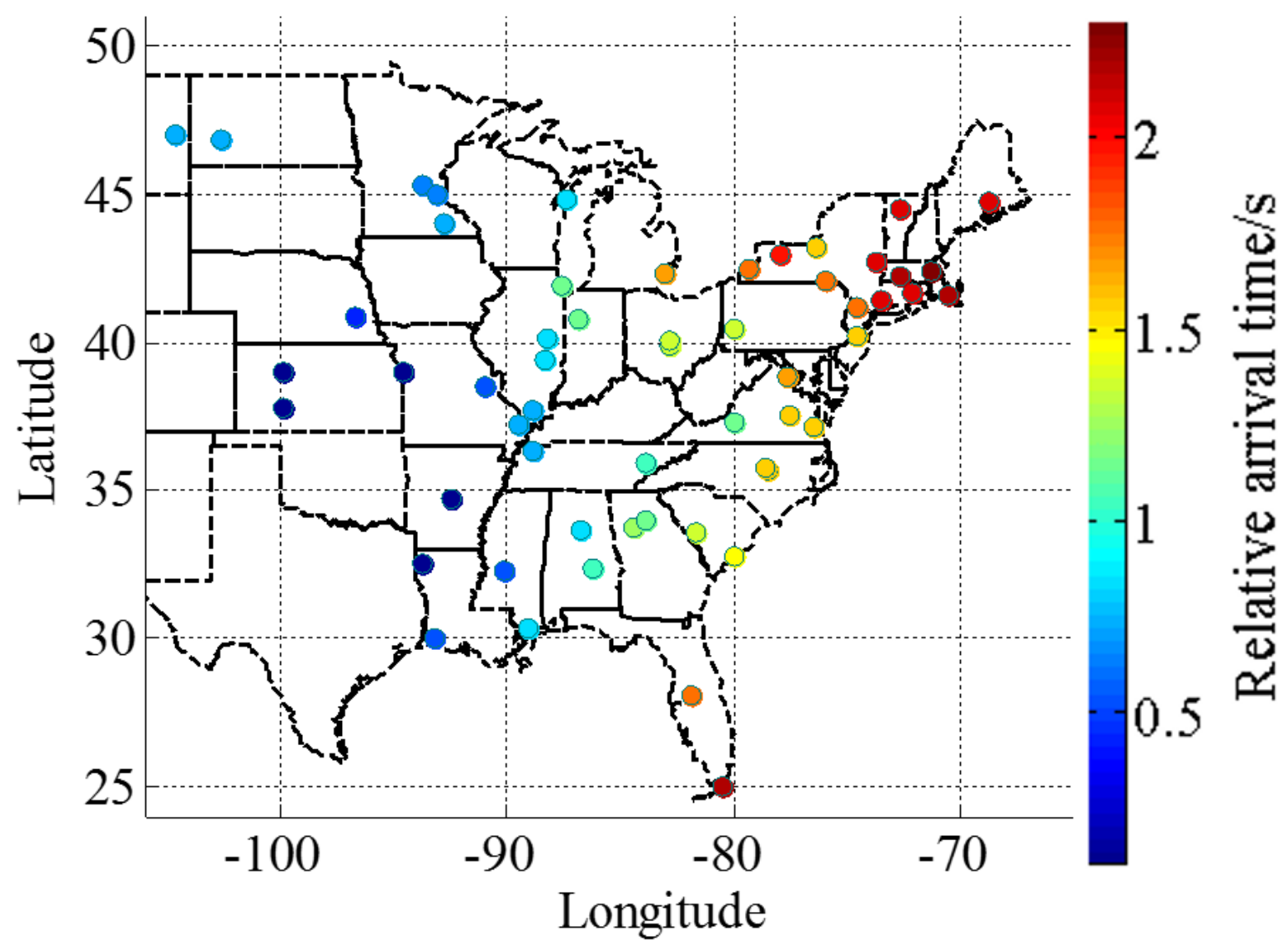

***Fig. 8.*** *Relative arrival time of FDRs at different locations*

### *3.2. Delaunay triangulation and bicubic 2D interpolation*

Intuitively, the true disturbance location would be in an area that has the minimum relative arrival time. Since the propagation speed is unknown and it may vary with the power grid status, state-of-the-art methods are highly depended on parameters and could not give a unique solution on the location or the start time of a disturbance. In contrast, this method combines Delaunay triangulation and bicubic 2D interpolation to calculate the disturbance location. The proposed approach is parameter-insensitive, robust against parameter errors. Triangulation of FDR locations. Delaunay triangulation partition the area by triangles using existing FDR locations in a nearest neighbour manner, ensuring that no FDR is within the circumcircle of a triangle formed by other three FDRs. In this method, the indicator of whether $\mathrm{FDR}_i$ is within the triangle formed by the location of another three FDRs ($\mathrm{FDR}_A$, $\mathrm{FDR}_B$, $\mathrm{FDR}_C$) is $M_{i-A,B,C}$ as shown in (2). This value should be positive for points lying inside the circumcircles when the FDRs at A,B,C are sorted in counter-clockwise. As Delaunay triangulation in a 2D space is a frequently-performed routine, the Association for Computing Machinery (ACM) archives a standard algorithm (Algorithm 872) [25].

$$M_{i-A,B,C} = \begin{vmatrix} lon_A - lon_i & lat_A - lat_i & (lon_A^2 - lon_i^2) + (lat_A^2 - lat_i^2) \\ lon_B - lon_i & lat_B - lat_i & (lon_B^2 - lon_i^2) + (lat_B^2 - lat_i^2) \\ lon_C - lon_i & lat_C - lat_i & (lon_C^2 - lon_i^2) + (lat_C^2 - lat_i^2) \end{vmatrix} \quad (2)$$

Fig. 9 shows the spatial Delaunay triangulation of FDR locations in the EI power grid. Delaunay triangulation minimizes the maximum angle of all triangulations that connect three FDR locations, allowing the reconstruction of the responses time at locations that have no FDR installed.

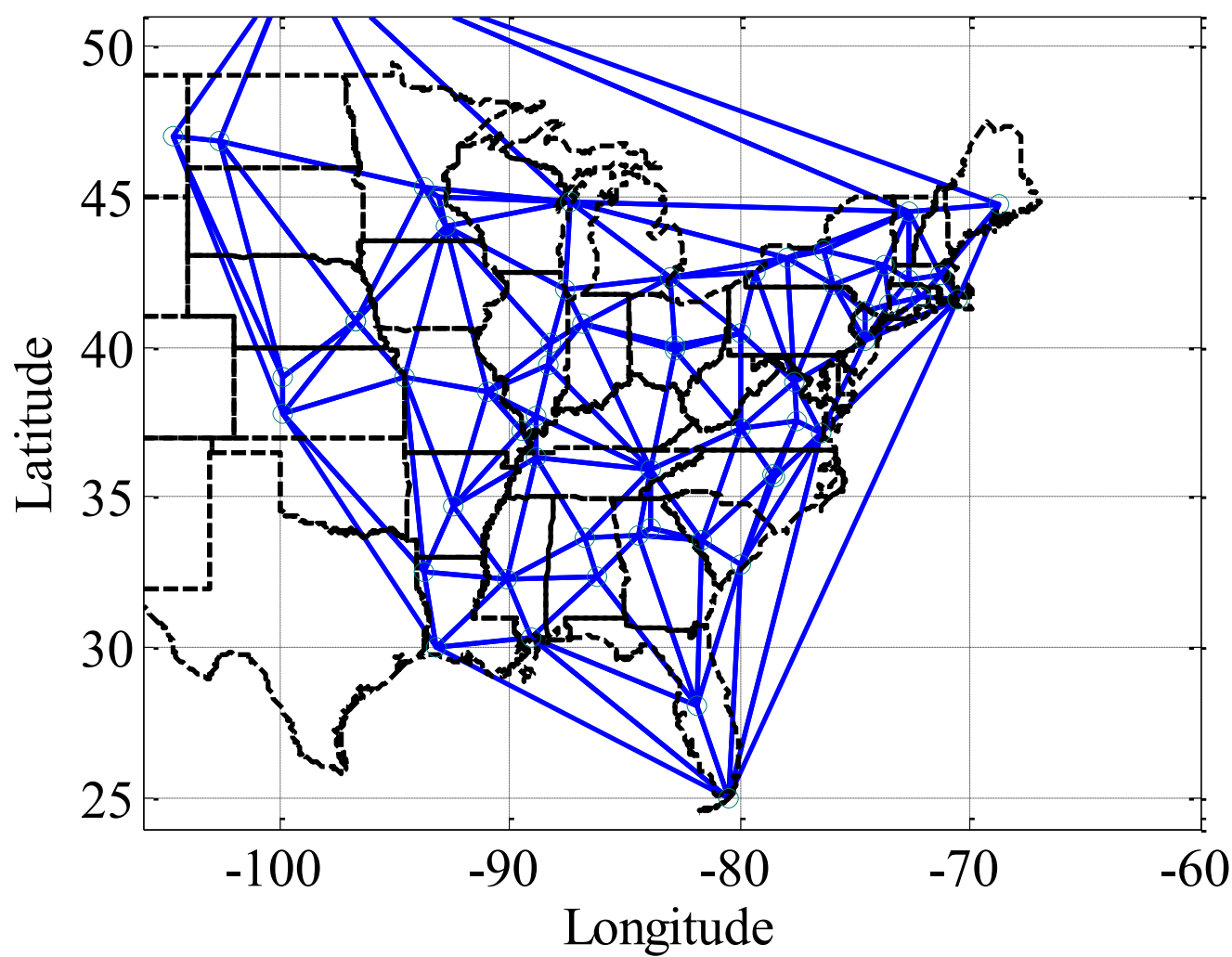

***Fig. 9.*** *Delaunay triangulation of FDR locations in the U.S. Eastern Interconnection*

After Delaunay triangulation, this method interpolates the FDR response time using bicubic 2D interpolation for each triangle. For the interpolated arrival time for each triangle has the following form.

$$T(lon, lat) = \sum_{u=0}^{5}\left(\sum_{v=0}^{5-u} a_{u,v} lon^{u} lat^{v}\right) \tag{3}$$

where $a_{u,v}$ are the parameters of the polynomial calculated using the triangle-based surface fitting method described in [26], improved from its previous version [26, 27]. An implementation of this algorithm is ACM Algorithm 761 as documented in [28, 29].

Fig. 10 shows the contour map of the bicubic 2D interpolation result. Bicubic 2D interpolation computes a two-dimension cubic function to fit the triangulated response time at the scattered points. The blue areas in Fig. 10 show the locations with smaller response time, indicating locations near the disturbance, whereas the red areas represent significant latency in response due to wave propagation delay.

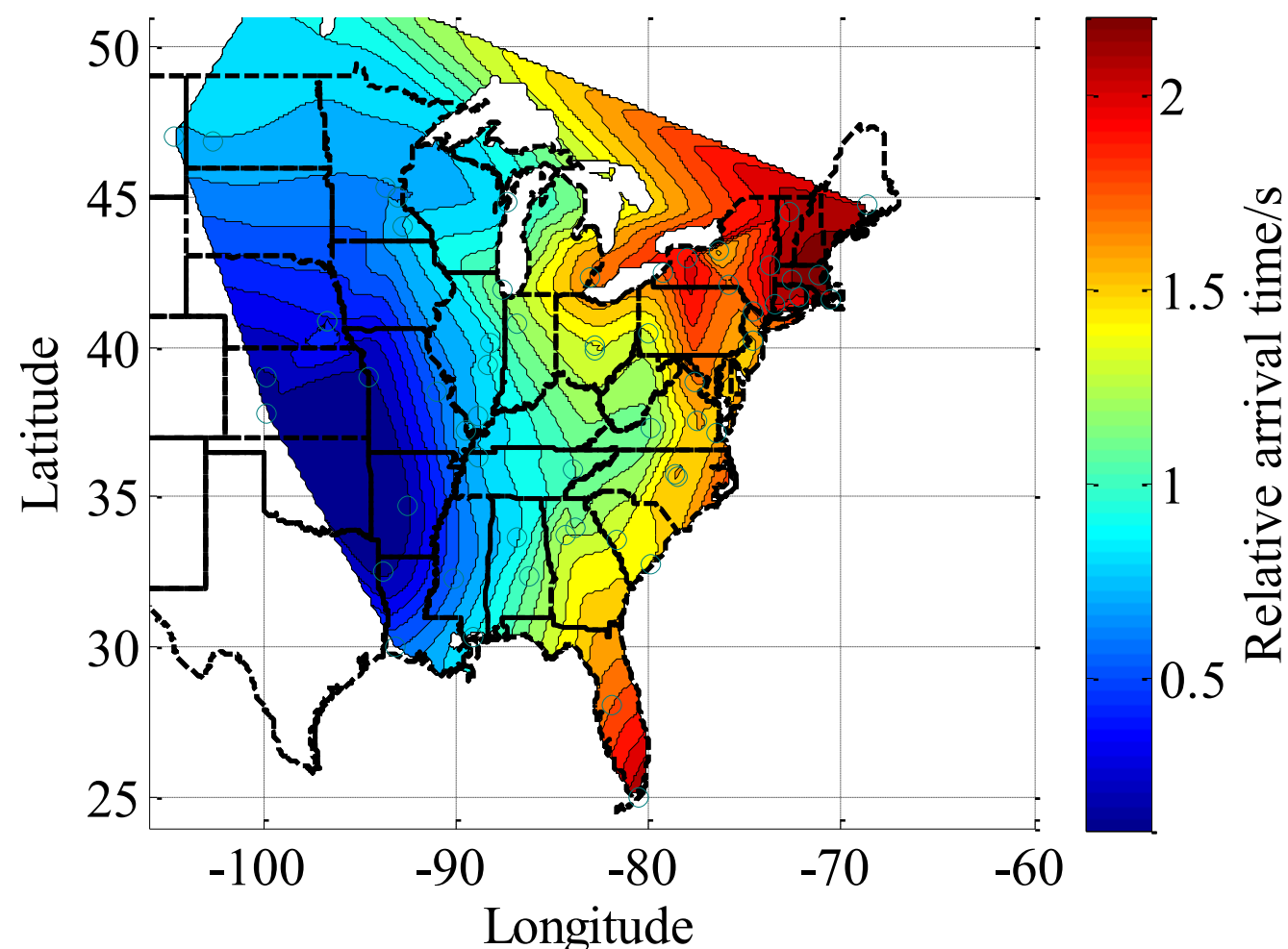

***Fig. 10.*** *The contour map of the time of ROCOF passing a threshold for all FDRs*

*3.3. Pinpointing event location and calculating the event start time*

The method then scans the mesh grid to look for the point that has the global minimum response time, which is designated as the estimated disturbance location, as shown in Fig. 11. For this case, the computational time consumption to find the disturbance location is 0.461 second using a desktop with a 3.2 GHz CPU. Fig. 12 shows a comparison on the actual and estimated disturbance location. The red, blue, and white dots denote the actual location, the estimated location based on proposed method, the estimated location based on the method in [30]. The distance between the actual and estimated disturbance location using the proposed method is 15.8 miles, while the value is 100.5 miles using the method in [30]. This result indicates the small error of the proposed method.

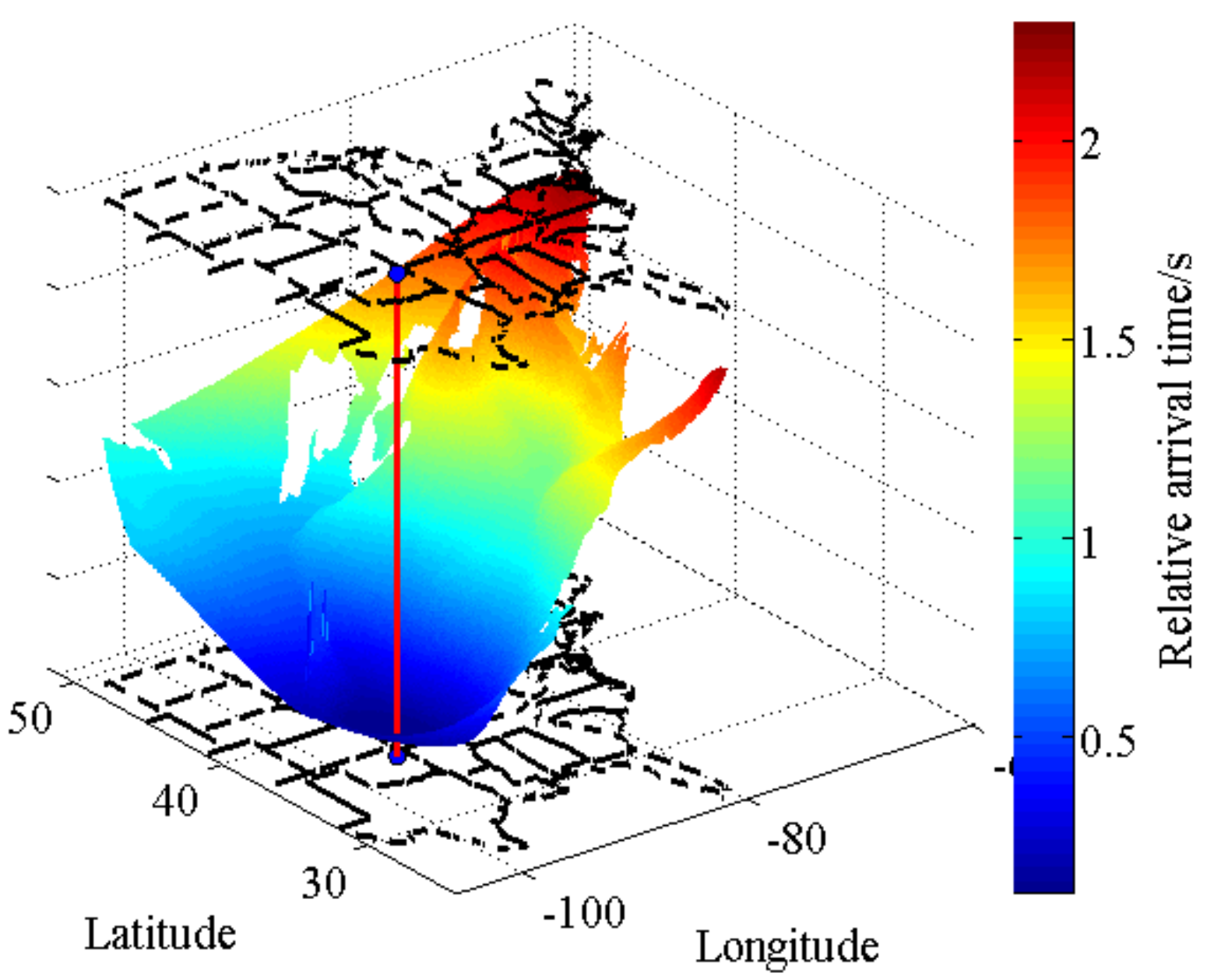

***Fig. 11***. *Pinpointing the event location on the contour map*

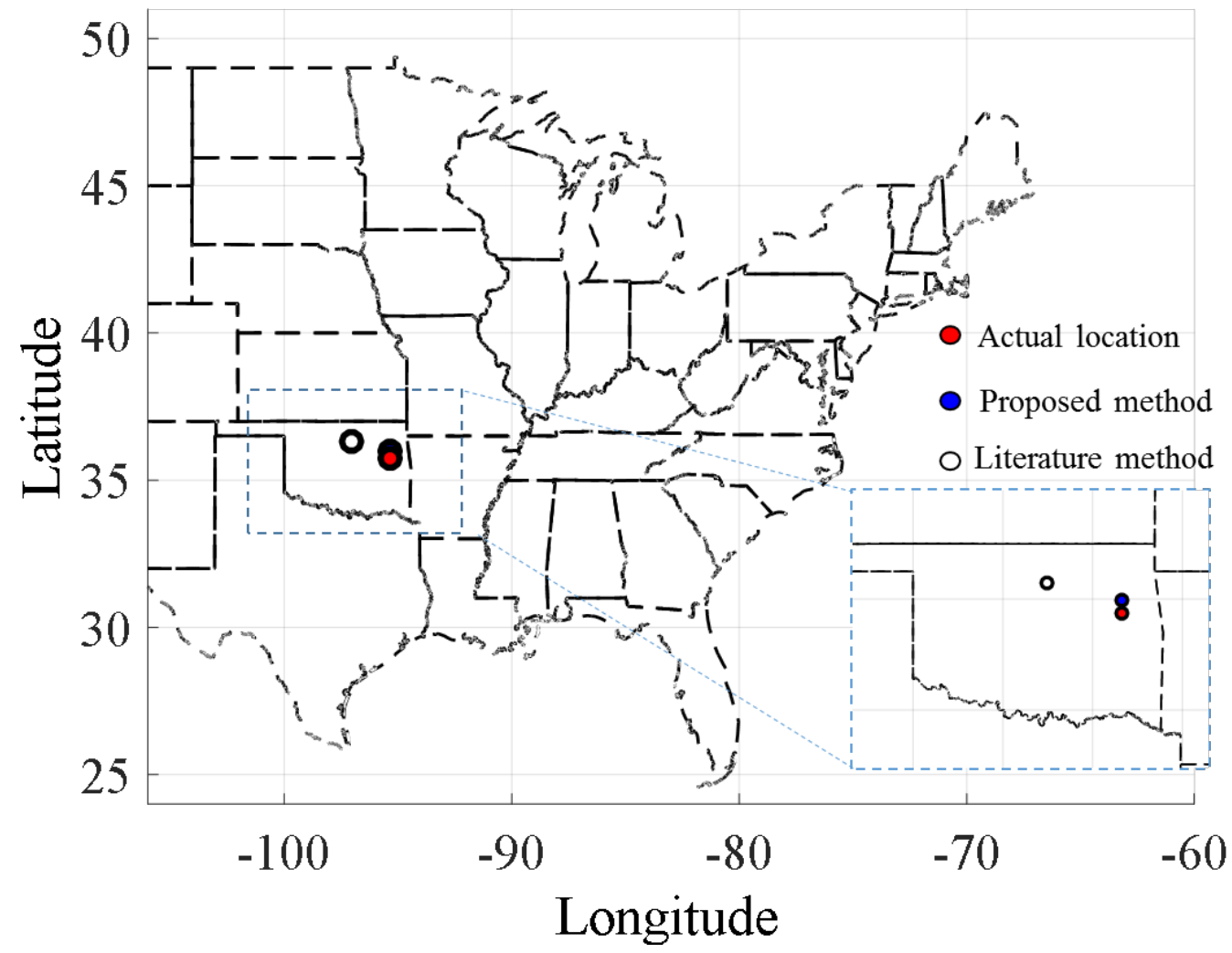

***Fig. 12.*** *Estimated and actual disturbance location*

Assuming the obtained disturbance location estimation is denoted by $(lon_{event}, lat_{event})$, the response time at this location is denoted by $t_{min}$. Then the actual disturbance start time is estimated as $t_{event} = t_{min} + t_R$, where $t_R$ is the common time reference for calculating the relative arrival time for all FDRs.

### *3.4. Data Validation*

FDR (or PMU) data may include bad data with some wrong timestamps due to GPS loss, clock error and leap second issues. Therefore, after obtaining the estimated event location and estimated event start time, it is necessary to double check the quality the credibility of the relative arrival time. This validation applies a linear regression method for data validation. The distance between the FDR location and the estimated event location is denoted as

$$D_{(FDR_n, event)} = Distance\{(lon_{FDR_n}, lat_{FDR_n}), (lon_{event}, lat_{event})\} \tag{4}$$

The measured propagation time is calculated as

$$\Delta t_{(FDR_n, event)} = t_{FDR_n} - t_{event} \tag{5}$$

Then the distance and the measured propagation time for each FDR: $\{D_{(FDR_n, event)}, \Delta t_{(FDR_n, event)}\}\big|_{n=1,2,..n}$ are checked using linear regression, assuming the propagation speed is constant, which means the propagation response time delay is proportional to the distance between the FDR location and estimated event location. A threshold is used to find the outliers of the measurements:

$$\left|\Delta t_{(FDR_n, event)} - \Delta t'_{(FDR_n, event)}\right| > \delta_t \tag{6}$$

where $\Delta t'_{(FDR_n, event)}$ is the propagation time from the estimated event location to FDR *n* obtained from linear regression; $\delta_t$ is the tolerance of the deviation between the measurement and the linear regression result. As a typical practice in identifying outliers in linear regression, $\delta_t$ is selected as 1.5 times of the interquartile range of near $\Delta t'_{(FDR_n, event)}$. If any FDR is found to be an outlier, then this problematic FDR will be reported to the operator and its data will be deleted before recalculating the event location using Delaunay triangulation and interpolation. An example of the measurement that has the time stamp issue is shown in Fig. 13. It can be seen that the Michigan State has a measurement with wrong timestamp, as marked by the black arrow.

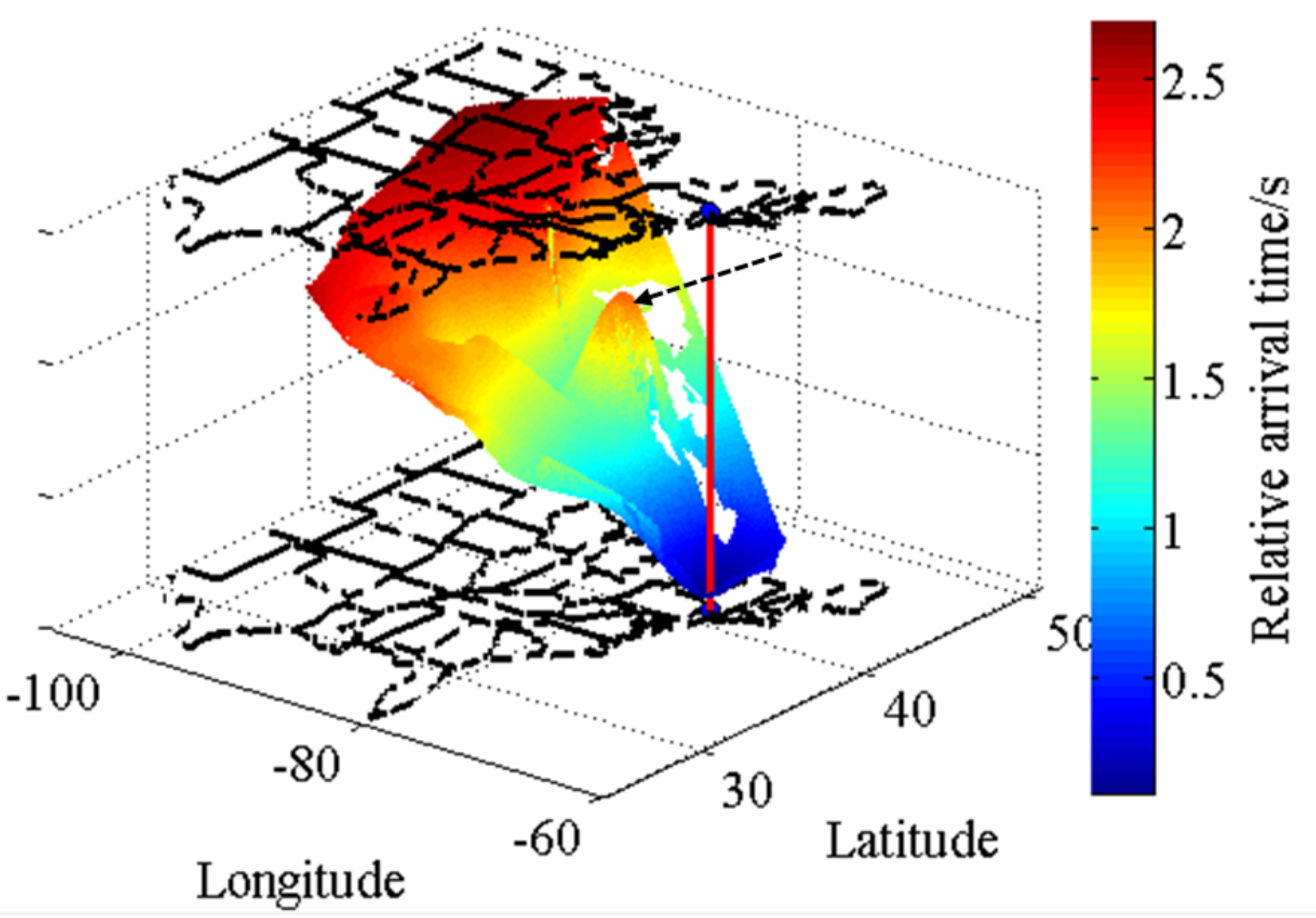

***Fig. 13.*** *Detecting measurement with a time stamp issue in Michigan, U.S.*

To test the robustness of the proposed method, eight events happened during the period from August, 2013 to June, 2015 were tested using the proposed algorithm. The average distance between the estimated location and actual event location is 19.3 miles with a standard deviation of 9.8 miles. Because the maximum angle in the triangle constructed on grid edges will have be larger due to the lack of FDRs on one side, the surface fitting result has relatively larger errors. Therefore, relatively larger errors are often seen for cases in which the actual disturbance location is near the grid edge.

### *3.5. Electromechanical Wave Propagation Speed Distribution Calculation*

Base on the interpolated arrival time distribution $T(lon, lat)$ from validated measurements, the propagation speed can be calculated by the following steps:

1) Calculate the local gradient of the arrival time for each location and for both the longitude and latitude direction, denoted as $\partial T(lon_i, lat_j)/\partial lon$ and $\partial T(lon_i, lat_j)/\partial lat$, respectively.
2) Rescale the directional gradients at each point based on the actual per unit distance in the longitude and latitude side, respectively.

$$\begin{cases} \partial T(lon_i, lat_j)/\partial d_{lon} = \left(\partial T(lon_i, lat_j)/\partial lon\right)/UniD_{lon} \\ \partial T(lon_i, lat_j)/\partial d_{lat} = \left(\partial T(lon_i, lat_j)/\partial lat\right)/UniD_{lat} \end{cases} \quad (7)$$

3) Obtain the local composite gradient of wave propagation delay based on the two directional gradients.

$$\partial T(lon_i, lat_j)/\partial d = \sqrt{\left(\partial T(lon_i, lat_j)/\partial d_{lon}\right)^2 + \left(\partial T(lon_i, lat_j)/\partial d_{lat}\right)^2} \quad (8)$$

4) Inverse the composite gradient to obtain the local propagation speed for location $(lon_i, lat_j)$.

$$v_{i,j} = 1/\left(\partial T(lon_i, lat_j)/\partial d\right) \quad (9)$$

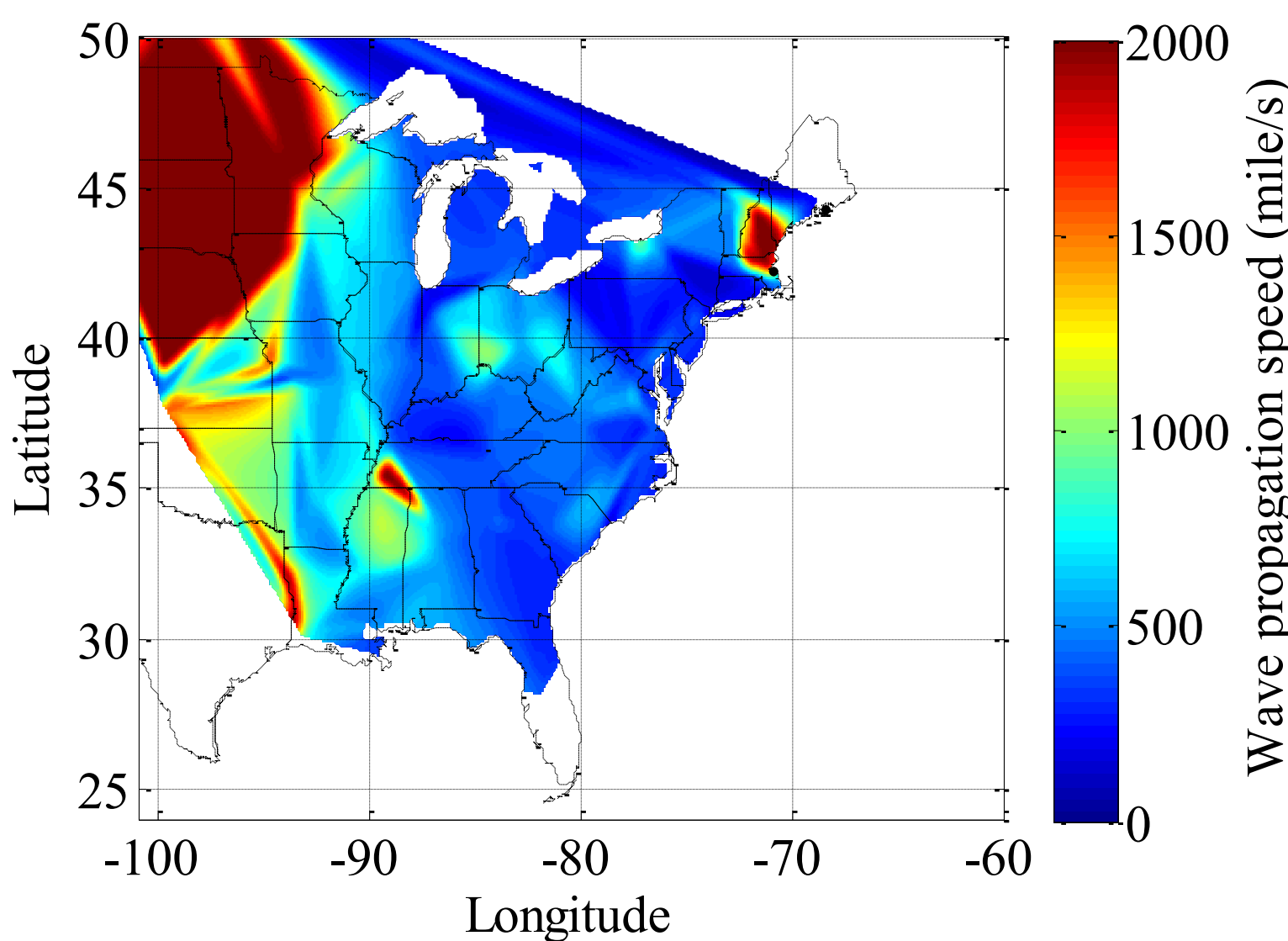

***Fig. 14.*** *Wave propagation speed distribution*

Fig. 14 shows the propagation speed distribution. The range of speed distribution (300-800 miles/s) in the east of EI is consistent with sample values observed by utilities in this area [31]. It shows that the propagation speed is higher in the western and lower in the eastern EI. This difference is because the eastern EI has more generation and load, thus having larger inertia than the western part. The higher inertia makes the eastern EI more robust to frequency fluctuations and slows down electromechanical wave propagation. The central EI has a relatively low propagation speed for the same reason. On the contrary, the western EI has less inertia and faster electromechanical wave propagation. Since the proposed method is based on electromechanical wave propagation. The method will work for large load shedding, line trip and faults, as long as the propagation of electromechanical wave propagation can be observed in measurements.

## 4. Conclusion

This paper demonstrates a new disturbance location determination method implemented on a wide-area frequency monitoring network — FNET/GridEye. Without requiring a pre-determined propagation speed value, the proposed method can accurately pinpoint the disturbance location in the power grid. In addition, this method is robust to timestamp-shifting and measurement error. Based on this method, the real-time distribution of electromechanical wave propagation speed can also be calculated. The proposed method has the generality to be implemented in other WAMSs.